# ELECTROKINETIC PUMPING AND ENERGY CONVERSION AT NANOSCALES

Chirodeep Bakli and Suman Chakraborty*

Department of Mechanical Engineering, Indian Institute of Technology Kharagpur, Kharagpur 721302, INDIA

This paper gives a lucid and integrated description of the works " Nonlinear Amplification in Electrokinetic Pumping in Nanochannels in the Presence of Hydrophobic Interactions [1]" published in electrophoresis. People interested to perform associated work either numerically or experimentally may find this helpful.

## Abstract

The integration of the coupling effects of intrinsic wettability and surface charge in a nanochannel can cause non-intuitive behavior in the electrokinetic energy conversion processes. We demonstrate that in a nanofluidic device the energy conversion efficiencies may get amplified with an increase in surface charge density, not perpetually, but only over a narrow regime of low surface charges, and may get significantly attenuated to reach a plateau beyond a threshold surface charging condition. This results from the complex interplay between fluid structuration and ionic transport within a charged interfacial layer. We explain the corresponding findings from our molecular dynamics simulations with the aid of a simple modified continuum based theory. We attribute our findings to the four-way integration of surface charge, interfacial slip, ionic transport, and the water molecule structuration. The consequent complex non-linear nature of the energy transfer characteristics may bear far-ranging scientific and technological implications towards design, synthesis and operation of nano-batteries which can supply power at scales of the range molecular dimensions.

## Introduction

Fluid as a carrier of energy in nanopores and nanochannels can be utilized to generate nanowatts of power which through a parallel tapping using a large array can work as an effective alternate energy source. Tapping the streaming potential in such a system would lead to conversion of the mechanical energy from flow to electrokinetic energy. Electrokinetic energy conversion (EKEC) processes have traditionally been found to yield extremely low efficiencies in the experimental domain [2,3], irrespective of theoretical methods to augment the efficiency. The existing theoretical analyses happens to miss the interdependence of the presumed slip lengths $(l_s)$ and the surface charging conditions [4–8]. In addition, they have not addressed the possibility of existence of any non-trivial coupling of substrate charging (or salt concentration), the near-wall structuration of water molecules, and the dynamics of interfacial slip over molecular scales, which could result in a complex dependence of the streaming current and streaming potential on the surface charge, and influence the energy conversion characteristics of a nanofluidic device in a rather profound manner [9–12].

## Simulation Details

We perform MD simulations to assess water density distribution $(\rho)$, slip length $(l_s)$, streaming current $(I_s)$, and streaming potential $(E_s)$, as a combined consequence of the driving force (expressed in terms of an equivalent acceleration, $g$) as well as the intrinsic wettability (expressed in terms of the static contact angle $\theta_s$) and the surface charge density $(\Sigma)$ of a nanofluidic channel. The channel walls consist of four layers of atoms arranged in FCC lattice in <100> plane, as shown in Fig.1. To simulate wall charging, we follow two different approaches. First, we evenly distribute the charge amongst the wall atoms, and secondly the same amount of charge is distributed as unit charges on randomly selected wall atoms. The results obtained in both the simulations do not show any appreciable difference and thus any method could be interchangeably used to obtain the simulation results. A typical unit cell chosen for simulation has dimensions $25 \times 22$ (lateral dimensions)$\times 16$ (y $\Rightarrow$ height) units. Periodic boundary conditions along *x* and *y* directions ensure effectively infinite dimensions in axial and transverse directions.

Length units are normalized using the Lennard-Jones (LJ) parameter $\sigma$ for the wall atoms. The wall atoms are packed with a density of $\rho\sigma^3 =1$. The number of water molecules in each simulation cell is adjudged from the density of water at 300K. The wall atoms are tethered to springs having a spring constant high enough to prevent the vibrations from exceeding the threshold value [13]. The springs make the wall flexible, as rigid walls introduce artifacts in obtaining slip lengths [14]. The Simple Point Charge/Extended (SPC/E) model is used to define the water molecules and their interactions [15]. The wall particles interact via LJ potentials parameterized by $\sigma$ and $\varepsilon$. The ionic species consist of monovalent Na$^+$ and Cl$^-$. The ions interact with the system via LJ potential and also by Coulomb forces. The $\sigma$ value of the wall particles is taken to be equal to that of the SPC/E water. The $\varepsilon$ values are varied according to the intrinsic wettability of the substrate. The contact angle subtended by the salt solution on the channel walls is obtained by allowing a drop to settle on the substrate in a separate set of equilibrium MD simulations. The salt concentration of the solution is varied from 7mM to 6M, which is simultaneously accompanied by a variation of the surface charge between 0.003 C/m$^2$ to 0.8 C/m$^{2..}$. The flexible wall atoms are thermostatted using Nose-Hoover thermostat and the solution bulk temperature is maintained by the water molecules dissipating through the flexible walls. The hetero-nuclear interactions are defined using Lorentz-Berthelot mixing rule [16]. The system is energy minimized and equilibrated for 1000 time units ($10^6$ time steps). The simulation is run for 5000 time steps (integrated using leap frog algorithm with step size of 0.001 units). The time scale normalized using $\tau = \sqrt{\dfrac{m\sigma^2}{48k_BT}}$, where $m$ is the mass of water molecule, $k_B$ is the Boltzmann constant, and $T$ is the absolute temperature.

The channel walls consist of four layers of atoms arranged in FCC lattice in <100> plane as shown in fig.1. A typical unit cell chosen for simulation has dimensions $25\times 22$ (lateral dimensions)$\times 16$ (2H $\Rightarrow$ height along y) units, where length units are normalized using the Lennard-Jones (LJ) parameter $\sigma$ for the wall atoms. The salt concentration of the solution is varied from $7\,mM$ to $6\,M$, which is simultaneously accompanied by a variation of the surface charge between $0.003\,C/m^2$ to $0.8\,C/m^2$. In order to measure the slip length, we plot the molecular velocity profile in the channel averaged over time and simulations [17–19]. The continuum prediction of a parabolic profile is superimposed on the above profile. Fitting the same with the Navier boundary slip condition $\left( l_s = \dfrac{u_{wall}}{(\partial u / \partial y)|_{wall}} \right)$, the slip lengths for each of the simulation cases are obtained.

To measure the streaming current and streaming potential, we average out results from several different (typically, of the order of 10 to 100 numbers) simulations running over a $8000\tau$ time frame. These simulations are preceded by energy minimization and equilibrium simulations. For streaming current measurement, we directly use the ionic velocity from the trajectory data. The summation of this data for individual ions in a cross-section averaged over the simulation time gives the streaming current. Determination of streaming potential is not quite straightforward given the periodic boundary conditions in the flow direction. We systematically vary an artificially applied back electric potential, within the unit cell, till the net current through the channel is zero. Through this hit and trial method we obtain the streaming potential value which in magnitude is equal to the back potential value at zero current. We express the EKEC efficiency $(\eta)$ in terms of the ratio of the output electrical energy $(I_s E_s)$ and the input mechanical energy $\left( Q \dfrac{\Delta P}{L} \right)$, where $Q$ is the flow rate through the channel under a pressure gradient of $\left( \dfrac{\Delta P}{L} \right)$, $L$ being the axial length under concern, so that

$$\eta = \dfrac{I_s E_s}{Q \dfrac{\Delta P}{L}} \qquad 0.1$$

This expression gives the open circuit efficiency for the device. It should be noted that the actual efficiency that is tapped in an electrical circuit is a fraction of open circuit efficiency and its magnitude would depend on the electrical/electronic design of the circuit.

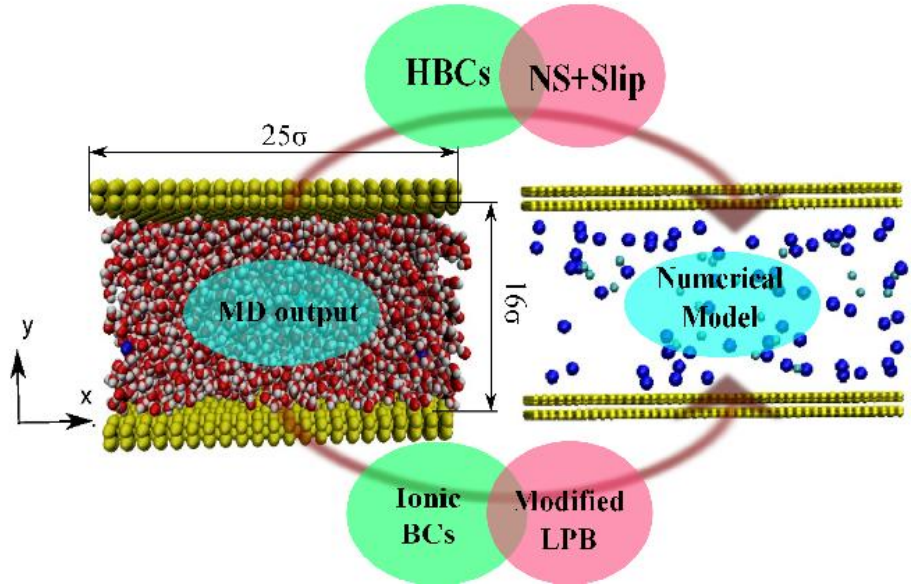

**Figure 1** Typical MD simulation domain. Water with ions dissolved in it is sandwiched between negatively charged walls. The ion distribution at a snapshot after equilibration depicts the breakdown of assumptions of the Boltzmann distribution and the continuum EDL theories. The cyan particles are the negatively charged co-ions and the blue particles are positive counter-ions. The representative unit cell in the right depicts the numerical model that approximates the fluid to be continuum. The ionic contributions and charge boundary conditions obtained from the modified Langevin-Poisson-Boltzmann (LPB) model, coupled with slip-based hydrodynamic boundary conditions (HBCs) for the Navier-Stokes (NS) equation, where the slip lengths are obtained from MD simulations, are used for the pseudo-continuum simulations.

**Pseudo-continuum description**

Because of the complex interconnections between the concerned physical parameters, such as the near-wall density fluctuations, $\Sigma$, and $l_s$, the standard continuum models of electrokinetics may turn out to be insufficient in capturing the correct trends in the energy conversion characteristics and modifications in the standard Poisson-Boltzmann formulation is required to capture the complete physics [20,21]. As an alternative, here we develop a novel pseudo-continuum description to capture the underlying effects. For simplicity in description, we consider a z:z symmetric electrolyte being pumped by a driving pressure gradient $-\nabla P$ through a slit-type nanochannel. Considering the resulting velocities of the positive and the negative ions to be combined consequences of the fluid velocity ($\vec{u}$) and electro-migration velocity, one may write $u^\pm = u \pm \dfrac{zeE_s}{f^\pm}$, where $f^\pm$ represents friction factors of the respective ionic species. With a simplified assumption of friction factors depending on the bulk ionic concentration (*CRC Handbook of Chemistry and Physics*), we can write: $f^{\pm} = f = \frac{2n_0 z^2 e^2}{\lambda}$, where $n_0$ is the bulk ionic concentration and $\lambda$ is the ionic conductivity. The net ionic current through any section may be obtained as:

$$I = \int_A ze(n^+ u^+ - n^- u^-) dA = ze \int_A (n^+ - n^-) u dA + \frac{z^2 e^2 E_s}{f} \int_A (n^+ - n^-) u dA \qquad 0.2$$

Under electroneutral conditions, $I = 0$. However, this does not lead to an immediate explicit determination of $E_s$, because of its inter-dependence, in turn, with the velocity field. The velocity field may be obtained by solving the momentum equation in the Stokes flow regime:

$$\mu \nabla^2 u - \nabla P + \rho_e E_s = 0 \qquad 0.3$$

, where $\rho_e$ is the volumetric charge density and is the integral of the local volumetric charge density $(\rho_e)$. In essence, we augment the standard analytical paradigm of streaming potential description in nanochannels by three distinct considerations. First, we express the local volumetric charge density $(\rho_e)$ as

$$\rho_e(y) = ze(n^+(y) - n^-(y)) - \frac{dPo(y)}{dy} \qquad 0.4$$

, where $n^+$ and $n^-$ are the number densities of the positive and negative ions in a $z:z$ symmetric electrolyte solution, $e$ is the protonic charge, and $y$ is the wall-normal direction. The term $\frac{dPo}{dy}$ takes into account the polarization effects due to orientational ordering of water [23–25], and essentially considers that the permittivity of an electrolyte close to a charged surface decreases on account of an enhanced orientational ordering of water diploes, as well as due to finite sizes of the ions and diploes.

Second, we take into account the fact that the solvent density distribution $\rho(y)$ not only alters the velocity field, but also alters the local number densities of the positive and the negative ions. Accordingly, one may write:

$$n^{\pm} = n_0 \exp{(\mp ze\psi - \psi_{ext})}/{k_B T} \qquad 0.5$$

where $\psi$ is the potential distribution within the EDL and

$$\psi_{ext} = -k_B T \ln \frac{\rho(y)}{\rho_f} \qquad 0.6$$

represents the effect of solvent density ($\rho_f$ being the bulk liquid density) [12]. The near-wall solvent density variations as a function of substrate charge are depicted in the supporting information. Very similar to the effect of ionic size in narrow confinements [26], the solvent fluctuations can also induce variation in the electrokinetic behavior. These density values are used to obtain the modified potential distributions. We observe that at low values of surface charge density, the fluctuations in the density of fluid molecule is less, with the near-wall density peak value not much higher than the bulk density levels. As the surface charge density is increased, the fluctuations increase and also the near-wall density maxima. Finally, at very high surface charge densities, the density fluctuations start reducing again. Along with it, the near-wall maxima also reduce.

Third, we augment the coupling between the electrostatic and hydrodynamic model by expressing $l_s = l_s(\Sigma, \theta)$ for given driving pressure gradient and ionic concentration, where the exact functional dependence is obtained from MD data. A physical basis of this consideration lies in the fact that $l_s$ not only gets coupled with surface charges on account of the direct flow resistance to ions due to the coulombic interactions between free and bound charge, but is also implicitly related to the near-wall fluid density fluctuations. The dependence of $l_s$ on surface charge gets much more pronounced in case of polar solvents like water where the partial charges on molecules shows a remarkably complex behavior in the presence of surface charges. Here $\psi$

is the potential distribution within the EDL, which may be obtained by solving the Poisson equation:

$$\nabla \bullet (\varepsilon \nabla \psi) = -\rho_e \qquad 0.7$$

(where $\varepsilon$ is the permittivity of the solution), thereby closing the system of equations mentioned as above. The local variation of the permittivity of the solution is taken into account by considering water molecules as Langevin dipoles, consistent with the Langevin Poisson-Boltzmann model. Our present model, thus, captures the dielectric discontinuity at the interface [24,25] and incorporates a slip that is both hydrodynamic and electrical in nature. In practice, the solvent density fluctuation values are obtained from the MD simulations and are directly substituted in the above equation (noting that $\rho_e$ is now a function of solvent density fluctuation) which is then numerically integrated to obtain EDL potential field $\psi$. This, along with a determination of the velocity field, enables the determination of $E_s$. In this manner, we essentially establish a pseudo-continuum model for EKEC that co-operates a slip description that is simultaneously hydrodynamic and electrical in nature.

**Results and Discussions**

The near-wall solvent density variations as a function of substrate charge are depicted in fig. 2. These density values are used to obtain the modified density and potential distributions. We observe that at low values of surface charge density, the fluctuations in the density of fluid molecule is less, with the near-wall density peak value not much higher than the bulk density levels. As the surface charge density is increased, the fluctuations increase and also the near-wall density maxima. Finally, at very high surface charge densities, the density fluctuations start reducing again. Along with it, the near-wall maxima also reduce.

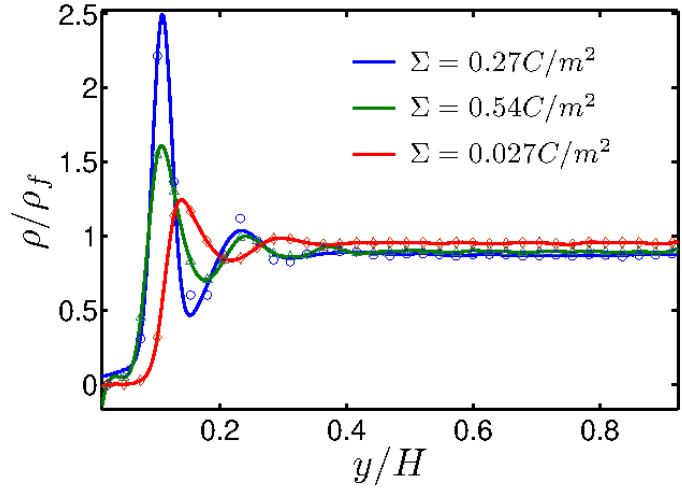

**Figure. 2** The normalized density fluctuations of water molecules across the channel of height $2H$. The variations in the degree of fluctuations and the height of the near-wall peak are non-monotonous, indicating the competition between the hydrodynamic and electrostatic interactions.

The slip length variation of NaCl solution over a charged surface of the nanochannel is depicted in fig. 3. We perform two different sets of simulations by varying the applied acceleration (in reduced units) and the modified Dukhin number ($\delta$) [27]. Here $\delta = N_w / N_c$, where $N_w$ and $N_c$ are the surface charge carriers and counterions in the fluid respectively. We first keep $\delta$ fixed and vary the applied pressure gradient, and then we vary $\delta$ and keep the pressure gradient constant. We observe, at low values of surface charge, the $l_s$ appears to be more or less constants in both the cases. With increasing values of $\Sigma$, the $l_s$ values are observed to decrease. For extremely high values of $\Sigma$, the $l_s$ almost becomes constant, and asymptotically converges to a value which corresponds to the no-slip condition.

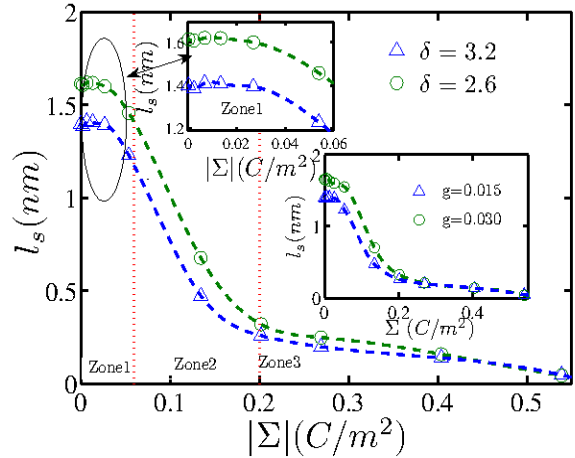

**Figure 3** Slip length variations with surface charge for different magnitudes of flow actuating force field and varying ratios of fixed surface charge and counterions (inset), for a substrate on which pure water has a contact angle of 120°. The effective slip at the wall remains constant at lower surface charge values. With increase of $\Sigma$, $l_s$ abruptly decreases before it plateaus at a constant value. $g$ is the acceleration provided by the applied force field (in normalized LJ units). With bare $\Sigma$ increasing at a rate greater than the increase in number of counterions, $l_s$ depletes faster, reaching the plateau at a lower $\Sigma$.

The peculiar and somewhat non-intuitive trend observed in fig. 3 can be rationalized as follows. At low values of surface charge, slip is a phenomenon grossly dictated by the intrinsic substrate wettability alone. As one increases the surface charge on the walls, the counterions are more favorably attracted towards the surface. These counterions agglomerate water molecules along with them in the vicinity of the surface, forming hydration shells. The coordination of water molecules around the ions, sticking on the surface, redevelops the Stern layer which was mobilized due to hydrophobic interactions in narrow confinements. With more water molecules sticking to the surface, the slip gradually decreases. At high values of $\Sigma$, a large number of counterions sticks to the surface and also coordinates water molecules around them. This leads to further reduction in $l_s$, and finally $l_s$ becomes a constant in an asymptotic limit. Further increase in surface charge does not trigger any more water agglomeration near the walls. Hence, $l_s$ becomes independent of the hydrodynamics with a layer constantly pinned at the charged wall.

This variation is coherent with the existing literature [28]. However, the possible interaction between this effect and the advective-electromigrative transport of ionic species in a nanofluidic channel and the corresponding alterations in EKEC characteristics have not been brought out yet. It is further important to note that for lower values of $\delta$, the slip $l_s$ decays more gradually with the increasing $\Sigma$. A lower surface charge is outweighed by the hydrodynamic drag which tries to continuously rearrange the water-ion agglomerations formed in the vicinity of the surface. However, beyond a certain threshold surface charge level, the $l_s$ characteristics merge. Beyond this $\Sigma$, $l_s$ becomes independent of salt concentration, applied shear or further rise in $\Sigma$. The $l_s$ variation, thus, can be divided into three regimes: A low $\Sigma$ zone where $l_s$ largely depends on the hydrodynamics, a transition zone where electrostatics gradually supersedes the hydrodynamic interactions to determine the molecular slip, and a high $\Sigma$ zone where $l_s$ becomes independent of hydrodynamics and solely depends on electrostatic interactions between water molecules and the free and bound charges.

We plot the EKEC efficiency $(\eta)$ as a function of the $\Sigma$, which is the central focus of this work. The primary inference that we derive from the plot is that altering the $\Sigma$ does not necessarily implicate a trivial variation in the energy conversion efficiency. We observe from fig. 4 that $\eta$ abruptly jumps to a peak with slight increment of $\Sigma$ in the low surface charge regimes, subsequently comes to a decline, and eventually reaches a plateau over high $\Sigma$ regimes. The initial rise of efficiency is due to the rise in $\Sigma$ of the walls and also the rise in fluid structuration which accompanies it. Over this regime, $l_s$ does not appreciably decrease and thus high conversion efficiencies result. The steep drop in efficiency results from the $l_s$ avalanche occurring at medium surface charge levels. At very high surface charge densities, the efficiency either remains constant or decreases slightly. Over this regime, $l_s$ has become constant, however the effect of increasing $\Sigma$ is countered by the reduction in fluid structuration resulting from ion overcrowding at the walls and the competition between these two forces results in the non-monotonic segments with a gradually decreasing nature.

Keeping $\theta_s$ unaltered, reduction in $\delta$ leads to visible efficiency increase in the lower surface charge regime. The peak efficiency values obtained are higher compared to the high $\delta$ counterpart. This increase can be attributed to the increase in number of free charge carriers in the solution with respect to the fixed charge. Also, we observe from fig. 3 that the $l_s$ values are higher for lower $\delta$ values; this increased slip can also be attributed to the increase in peak efficiency. With rise of surface charge, however, the efficiency values converge over a narrow regime specifically based on $\theta$. This trend can also be mapped back to fig. 3 where $l_s$ values are independent of $\delta$ at high values of bare surface potential.

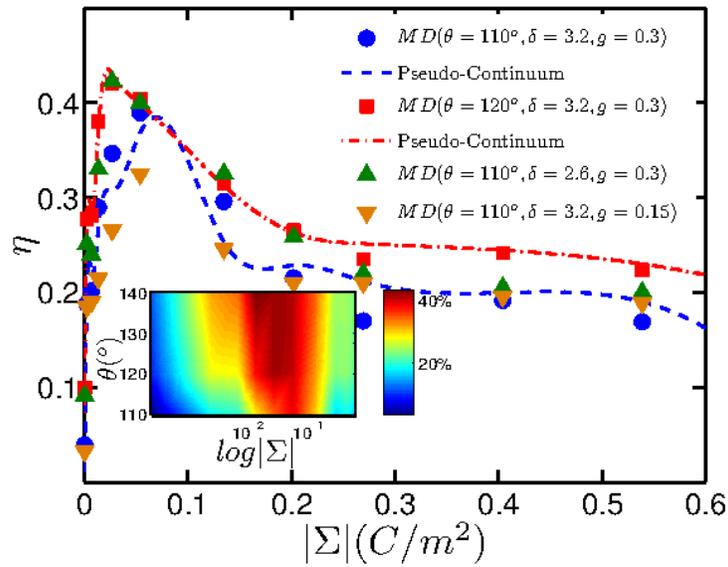

**Figure 4** Variation of EKEC efficiency with $\Sigma$. The efficiency first steeply increases, followed by a steep decrease and subsequently a virtual invariant nature. The variation of efficiency is depicted for different cases of surface wettability, shear and fixed charge-counter-ion ratio. Prediction from our pseudo-continuum description agrees qualitatively and quantitatively with the trend obtained from MD simulations. For clarity of plot, we show the pseudo-continuum description for two different $\theta$ s only. The efficiency emerges as a strong function of wettability and surface charge attaining maximum efficiencies (expressed as percentage) for intermediate $\Sigma$ values in for non-wettable surfaces as shown in the inset (where $\delta = 3.2$ and $g = 0.030$).

Variation of shear is also observed to affect the efficiency appreciably at lower values of surface potential. Decreasing $g$ leads to a dip in the peak efficiency. Eventually, for higher surface charge densities, the efficiency characteristics converge over a narrow regime, specific to the given $\theta$. Thus, we can infer that $\theta$ and $\Sigma$ are the two most important factors deciding the energy conversion efficiency of a nanofluidic device. The driving shear and $\delta$ values only affect the efficiency peak and can be fine tuned to obtain higher efficiencies, given the other factors are constrained by physical design parameters.

Interestingly, we also observe that the peak efficiency value occurs at a lower surface charge for a channel with higher contact angle value. This shifting of peak can be attributed to the change in water structuration influenced by salt concentrations over surfaces of various wettabilities. Salt solutions tend to induce greater density fluctuations when the surface is less wettable [18]. Thus, in channels of greater hydrophobicity, the maxima are achieved at lower concentrations and surface charges in a zone where hydrodynamics play a more pivotal role than electrostatics. The same argument applies to the cases of lower $\delta$ and lower driving acceleration. A lower value of $\delta$ would mean greater tendency of water to form the structures under lesser interaction of free and fixed charge, which explains the peak at a lower surface charge value. One may also observe that the width of the peak is greater, implying a greater range of surface charge enjoying higher efficiency values. This is again true for lower values of driving acceleration, which also indicate grater structuration and hence, wider peaks.

Proceeding further, we compare the pseudo-continuum predictions based on the model description outlined in the previous section with our MD simulation data. Since, wettability is the most important parameter which decides the gross energy conversion behavior; we choose the same parameter to analyze the validity of our simulations. The results (see fig. 4), indeed, agree both qualitatively and quantitatively. Such remarkable agreement between simplistic one-dimensional model predictions and MD simulation data may be realized only through the incorporation of surface charge density-dependent slip length and structuration effects in the pseudo-continuum formulation, borrowing concepts from MD considerations. Incidentally, $l_s$ as a sole function of $\theta$ would have predicted monotonously increasing efficiency characteristics, whereas non-inclusion of the structuration effects would result in omission of the efficiency peak shift effect that is observed in the various cases discussed above. Thus, the structuration not only increases the effective efficiency, but also enables one to tap the increased efficiency over a greater range of $\Sigma$ values. Hence, we observe that EKEC efficiency, which was traditionally believed to be an attribute of the substrate wettability and $\Sigma$ in a rather unrelated manner, in reality depends on an implicit coupling of the Σ and interfacial slip, as well as the near-wall structuration effects. These factors get combined to result in the conversion efficiency characteristics that cannot be rationalized through traditional continuum considerations alone.

How do our predictions compare with the results reported in prior literature on EKEC in nanochannels of comparable dimensions? In an effort to assess the same, we first set the slip lengths in our pseudo-continuum model to zero. The resulting predictions are able to replicate the results from recently reported Density Functional Theory (DFT) based studies [29]. Similarly, on substituting constant slip lengths, our pseudo-continuum model can reproduce the results from other studies which did not consider the complex electro-hydrodynamic coupling in the description of $l_s$ as realized through the MD studies reported here [30,31]. However, if this complex coupling is taken into consideration, the EKEC characteristics alter dramatically (see fig. 4).

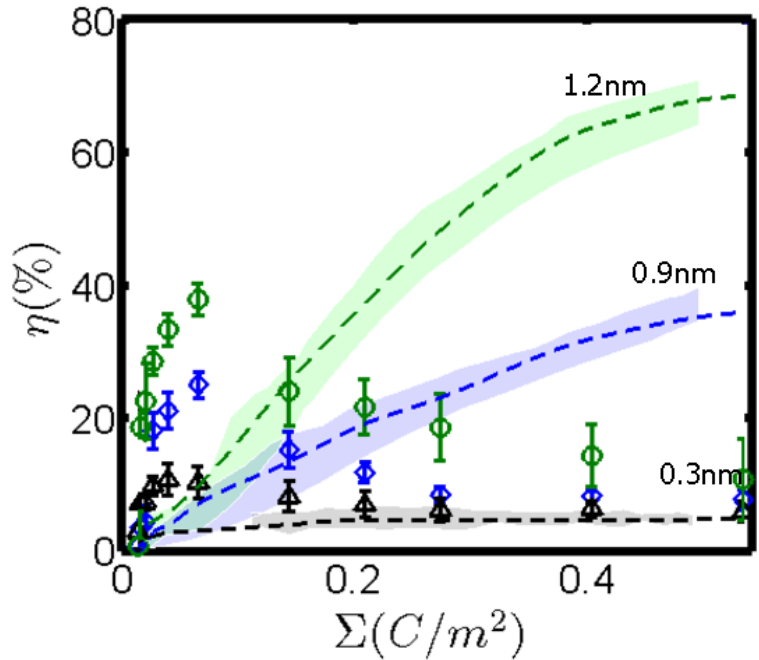

**Figure 5** EKEC efficiency variation with wall Σ for various sizes of counter-ions. The shaded plots represent the data obtained in [29]. The dashed lines show the predictions obtained from the pseudo-continuum analysis on setting the $l_s$ values to be zero for a substrate with contact angle of

$125^o$. The markers show the rsults obtained from MD simulations $\left(o\text{-1.2nm}, \lozenge\text{-0.9nm}, \square\text{-0.3nm}\right)$ which depicts the dramatic effect upon incorporation of slip lengths.

At low $\Sigma$ values, slip enhances the efficiency, whereas, at high $\Sigma$ the reducing $l_s$ causes deterioration of the device efficiency. Increase in counter-ion diameter proves even more detrimental to the EKEC efficiency, causing a more rapid decay in the same at higher values of $\Sigma$. Larger ionic particles can coordinate larger number of water molecules as compared to smaller ions. Such large hydrated structure drastically reduces slip and often leads to multilayer sticking. Since, in our analysis the co-ion is not changed ($Cl^-$) in any of the cases, the mobility of the co-ions remains unaffected and increased co-ion movement for higher surface charge leads to efficiency depletion for larger counter-ions.

**Conclusions**

This study demonstrates that efficiency augmentation in a nanofluidic energy conversion device is obtained for a narrow window of intermediate surface charge. This is in sharp contrast with the traditional theoretical predictions that may be obtained by employing surface charge-independent slip length based predictions of the conversion efficiency. We not only suggest a pseudo-continuum numerical model, but also compare predictions obtained from the same with the results independently obtained from MD simulations. We also compare our results with density functional theory (DFT) based analysis [29], showing the importance of the relation between surface charge and interfacial slip towards modulating the energy conversion characteristics of futuristic nanofluidic devices approaching molecular dimensions. With far-reaching practical implications, this paradigm opens up a new frontier concerning optimal functionalities of nanofluidic devices in terms of their energy conversion characteristics, instead of intuitively targeting for an illusive high-performance device bearing ultra-high surface charges. The key findings from this study are:

1) For a charged surface, slip lengths not only depend on the intrinsic surface wettability but also on the charge of the surface.

2) High surface charge density introduces more charge carriers in the fluid and prospectively increases the current. Slip also remobilizes the Stern layer, increasing the number of ions in the bulk. However, a combination of the two cannot be obtained for linear augmentation of current/efficiency.

3) The interaction of the surface charge, the ions and the polar water molecules leads to the pinning of water molecules and this reverses the effect of remobilized Stern layer.

4) This can effectively lead to the development of low surface charge based nanoscale energy conversion devices with practicable augmentations in the conversion efficiency, designed to exploit this complex interplay of the surface charge and the intrinsic substrate wettability characteristics.

To summarize, our investigations have revealed that involved interfacial phenomena may bear non-trivial implications on the energy conversion characteristics of nanofluidic devices, so that the EKEC efficiency may decline sharply to attain a plateau beyond a threshold increment in the surface charge, irrespective of the action of interfacial slip. Fundamental genesis of the underlying non-triviality lies in a complex coupling between the underlying electrostatics and hydrodynamics over interfacial scales.